\documentclass[12pt]{article}
\usepackage{latexsym}

\def\bs{\begin{subequations}}
\def\es{\end{subequations}}

\catcode`\@=11

\newtoks\@stequation
\def\subequations{\refstepcounter{equation}
  \edef\@savedequation{\the\c@equation}%
  \@stequation=\expandafter{\theequation}%   %only want \theequation
  \edef\@savedtheequation{\the\@stequation}% % expanded once
  \edef\oldtheequation{\theequation}%
  \setcounter{equation}{0}%
  \def\theequation{\oldtheequation\alph{equation}}}

\def\endsubequations{\setcounter{equation}{\@savedequation}%
  \@stequation=\expandafter{\@savedtheequation}%
  \edef\theequation{\the\@stequation}\global\@ignoretrue}

\makeatletter%  allow access to internal LaTeX commands
        \renewcommand{\theequation}{\thesection.\arabic{equation}}%
        \@addtoreset{equation}{section}%
\makeatother%  turn off access to internal LaTeX commands

\renewcommand{\thefootnote}{\fnsymbol{footnote}}

\begin{document}

\begin{titlepage}
Revised March 3, 2017
 
\begin{center}        \hfill   \\
            \hfill     \\
                                \hfill   \\
\vskip .25in

\begin{center}       
           
{\large \bf Revised Theory of Tachyons in General Relativity \\}
\vskip 0.54 cm

Charles Schwartz\footnote{E-mail: schwartz@physics.berkeley.edu}

{\em Department of Physics,
     University of California\\
     Berkeley, California 94720}
\vskip 0.3in

\end{center}

\vskip .3in

\begin{abstract}

A minus sign is inserted,  for good reason,  into the formula for the Energy-Momentum Tensor for tachyons. This leads to remarkable theoretical consequences and a plausible explanation for the phenomenon called Dark Energy in the cosmos. 

\end{abstract}
        
\end{center}
\end{titlepage}

\renewcommand{\thefootnote}{\arabic{footnote}}
\setcounter{footnote}{0}
\renewcommand{\thepage}{\arabic{page}}
\setcounter{page}{1}
\newpage 
\section{The Older Theory}

In a previous paper \cite{CS} I studied how a classical tachyon (faster-than-light particle), if such  a thing exists, would behave within the conventional theories of Special and General Relativity. If a particle follows a trajectory in space-time described by $\xi^\mu (\tau)$, then we write the following invariant, according to Special Relativity,
\begin{equation}
\dot{\xi}^\mu\;\dot{\xi}^\nu\; \eta_{\mu \nu} = \epsilon, \label{a1}
\end{equation}
where the dot signifies derivative with respect to $\tau$ and $\eta_{00} = +1,\; \eta_{11} = \eta_{22} = \eta_{33} = -1$. If we have $\epsilon = +1$, this is an ordinary particle, moving with speed $v < c$; and if $\epsilon = -1$, this is a tachyon, moving with speed $v > c$. A standard representation is (in units where $c=1$)
\begin{equation}
\dot{\xi}^\mu = (\gamma, \gamma \textbf{v}), \;\;\;\;\; \gamma = 1/\sqrt{|1-v^2|}.\label{a2}
\end{equation}

I then followed procedures familiar for ordinary particles and wrote down the Energy-Momentum tensor for any such point particle,
\begin{equation}
T^{\mu \nu} (x) = m \int d\tau\; \dot{\xi} ^\mu (\tau)\; \dot{\xi}^\nu (\tau)\; \delta^4 (x - \xi(\tau)).\label{a3}
\end{equation}
The first remark was that for very low energy tachyons, which have very large velocities, the space components of this tensor would be very large.
\begin{equation}
T^{i i} (x) = m \int d\tau\; (\gamma v_i)^2 \delta(t-\gamma \tau) \delta^3(\textbf{x} - \textbf{v}\gamma \tau) = m \gamma v_i^2 \delta^3 (\textbf{x} - \textbf{v} t).\label{a4}
\end{equation}

With this tensor as the source in Einstein's General Theory, I then calculated the gravitational fields produced by free moving tachyons and found that there would be a strong attractive force on other nearby tachyons moving parallel (or antiparallel). This led to a picture of rope-like structures of many tachyons that would produce strong gravitational fields in their vicinities. I thought that this structure might, if somehow attached to a galaxy, offer an explanation for the observed physical effects now attributed to Dark Matter: galactic containment and gravitational lensing. 

\section{The Revised Theory}

Simply put, there should be a minus sign in Equation (\ref{a3}) for a tachyon. 
\vskip 0.5 cm 

How I came to this idea is an amusing story; but here I shall just show the derivation.

We begin with the stationary principle for an arbitrary particle in the formalism of General Relativity, given a metric $g_{\mu \nu}(x)$ that can vary with spacetime coordinates $x$, replacing the flat-space metric $\eta_{\mu \nu}$.
\begin{equation}
A= \int d^4 x\; {\cal{L}} (x) , \;\;\; {\cal{L}} (x) = m \int d\tau\; \sqrt{\epsilon\; \dot{\xi}^\mu\; \dot{\xi}^\nu\; g_{\mu \nu} (\xi) }\; \delta^4 (x - \xi(\tau)),\label{b1}
\end{equation}
Here, as in Equation (\ref{a1}),  $\epsilon = +1$ denotes an ordinary particle ($v/c <1$) and $\epsilon = -1$ designates a tachyon ($v/c >1$). The epsilon must be placed under the square root sign so that this whole argument will be positive.  When we ask that the action ${\cal{A}}$ be stationary with respect to variation of the particle trajectories $\xi^\mu (\tau)$, we get the familiar geodesic equation, 

\begin{equation}
\ddot{\xi}^\nu + \Gamma ^\nu _{\alpha \beta} \; \dot{\xi}^\alpha \dot{\xi}^\beta = 0,\label{b2}
\end{equation}
where $\Gamma$ is the Christoffel symbol, involving derivatives of the metric $g_{\mu \nu}$; and we note that there is no epsilon appearing in this equation.

However, when we calculate the Energy-Momentum tensor for this Lagrangian density, either by Noether's theorem or by varying the  action (\ref{b1}) with respect to the metric $g_{\mu \nu}$,
we do get the epsilon out in front,
\begin{equation}
T^{\mu \nu} (x) =\epsilon \;m \int d\tau \; \dot{\xi}^\mu\; \dot{\xi}^\nu\; \delta^4 (x-\xi (\tau)).\label{b3}
\end{equation}

If we calculate the divergence of this $T^{\mu \nu}$, using (\ref{b2}) and (\ref{b3}), we get
\begin{equation}
\partial_\mu\; T^{\mu \nu} = - \Gamma^\nu _{\alpha \beta} T^{\alpha \beta},\label{b4}
\end{equation}
which is the usual result (without any epsilon) and we can then go on to put in the factor of $det(g)$ and  go to covariant derivatives for the full structure of Einstein's equation.

Given many particles, we can simply add their contributions to ${\cal{A}}$ , each one with its own mass and epsilon. They each obey their own geodesic equation (\ref{b2}). Similarly, they each contribute additively to $T^{\mu \nu}$ as in (\ref{b3}); but here each epsilon makes an important contribution.  There is, however, the alternative choice of putting a minus sign (the same epsilon) in front of the mass parameter for each tachyon in the action. This will return us to the original formula (\ref{a3}) for tachyons in the Energy-Momentum tensor. In this paper we wish to explore the first choice: keep all the contributions to the action as positive quantities.

With this new $\epsilon = -1$ factor entering the tachyons' Energy-Momentum tensor, we will have repulsive, not attractive, forces between colinear flows of tachyons. That interesting physical picture and mention of Dark Matter - given in my earlier paper - blows away.

But wait! There is something else big and new here. If we have lots of tachyons in the universe, they will all be flying about repulsed by one another but they still may contribute something important to the overall $T^{\mu \nu}$ as it is used to construct the Robertson-Walker Metric of the cosmos.  There will be NEGATIVE Pressure, and it may be large:. That is what the idea of Dark Energy was invented to provide.

\section{Neutrinos and Numbers}

Rather than imagine some unknown things to be tachyons, it is more conservative to suggest that some physical things, known but not completely known to us, may actually be tachyonic.  Neutrinos are the best candidate. They interact very weakly with ordinary matter or radiation and they are plentiful.

From various experimental results it is believed that neutrinos have a mass that is very small but not zero. An estimate is of the order of $m \sim 0.1 eV/c^2$.

From reigning cosmological models \cite{SW} it is thought that there is a vast sea of low energy neutrinos, called Cosmic Neutrino Background,  with an average density around $300/ cm^3$ and an average energy E of around $10^{-3 }eV$ per particle.

If these are actually tachyons, then we would say they have a typical velocity of $v/c \sim mc^2/E \sim 10^2$.

From the previous discussion and Equation (\ref{a4}), this would give a negative pressure amounting to about $ mc^2\; (v/c)$ times the density of such particles:  $P \sim 3,000 \;eV/cm^3$. This is of the correct order of magnitude to replace the current postulates of Dark Energy.

Obviously, there are a number of more detailed experimental and theoretical questions to be asked and answered before this hypothesis should be accepted as true. It will require others, more broadly experienced than I, to become engaged.

\section{Further Questions}
My older theory offered a possible explanation for Dark Matter; but with epsilon = -1, the force binding together flows of tachyons is gone.  We now envision a universe filled with a sea of tachyonic neutrinos, which repulse each other (gravitationally) and also repulse ordinary matter.
If galaxies form, as a fairly tight collection of ordinary matter, they might push a hole in that background sea of tachyon neutrinos; but then the surrounding sea would exert a "compressive" gravitational force on all that matter in the galaxy.  Could this be an alternative explanation for "Dark Matter"? This compressive gravitational force would keep the outer parts of the galaxy rotating at a faster velocity; and it could also be a major factor in gravitational lensing.

While this theory gives repulsive forces between tachyons and attractive forces between ordinary particles, other combinations appear strange. The gravitational force from tachyons repels ordinary particles while the gravitational force from ordinary particles attracts tachyons.  This appears to violate Newton's third law. That law comes from nonrelativistic theory and is often summarized as guaranteeing overall conservation of momentum. In our relativistic theory that conservation is held by the vanishing divergence of the energy-momentum tensor. Still, this is puzzling.

The numbers cited above describing the Cosmic Neutrino Background come from the application of classical theory of statistical mechanics. Is that correct for tachyons?  I don't know.

Everything in this paper has been formulated in terms of classical (but relativistic) theory of particles. What about quantum mechanics and quantum field theory?  I have written some about that \cite{CS2} but questions remain to be explored. The following section is a brief sortie into field theory.

\section{Complex Scalar Fields}

Let's look at a Lagrangian for a system that involves two types of particles, each described by a complex scalar field; and they interact. The first field will describe an ordinary particle and the second will describe a tachyon.
\begin{equation}
{\cal{L}} = (\partial^\mu \phi_1^* \partial_\mu \phi _1- m_1^2 \phi_1^* \phi_1) +
 \epsilon(\partial^\mu \phi_2^* \partial_\mu \phi_2 +m_2^2 \phi_2^* \phi_2)  - V(\phi_1 ,\phi_1^*, \phi_2, \phi_2 ^* ),
\end{equation}
where I have introduced a factor $\epsilon$ that may be chosen as +1 or -1.
The equations of motion are now,
\begin{equation}
\partial^\mu \partial_\mu \phi_1 + m_1^2 \phi_1 = - \frac{\partial V}{\partial \phi_1^*}, \;\;\;\;\;
\partial^\mu \partial_\mu \phi_2 - m_2^2 \phi_2 = - \epsilon \frac{\partial V}{\partial \phi_2^*}.
\end{equation}

We construct the energy-momentum tensor for the interacting system.
\begin{equation}
T^{\mu \nu} = (\partial^\mu \phi_1^* \partial^\nu \phi_1 + \partial^\nu \phi_1^* \partial^\mu \phi_1) + \epsilon'(\partial^\mu \phi_2^* \partial^\nu \phi_2 + \partial^\nu \phi_2^* \partial^\mu \phi_2) - \eta^{\mu \nu} {\cal{L}},
\end{equation}
and we find that $\partial_\mu T^{\mu \nu} = 0$ with only the conditions $\epsilon = \epsilon'$ and $V^* = V$. (This follows Noether's theorem.)

Conventional theoretical work would have led me to write the Lagrangian with $\epsilon = +1$. The new possibility here is the choice $\epsilon = -1$.   If we look at a space-space component of the tensor for a tachyon, dropping the interaction potential, we find,
\begin{equation}
T^{11} = \epsilon [ \;|\partial_0\phi|^2 + |\partial_1\phi|^2 -|\partial_2\phi|^2 - |\partial_3\phi|^2 + m^2 |\phi|^2].
\end{equation}
For a plane wave solution  this is just $2\epsilon |\partial_1\phi|^2$, which will give a negative pressure if we choose $\epsilon = -1$.

\vskip 0.5 cm
	
{\bf Acknowledgments}
I thank J. Ooms for assisting me with R.

\end{document}